# Field dependent magnetization of $BiFeO_3$ in ultrathin $La_{0.7}Sr_{0.3}MnO_3$/$BiFeO_3$ superlattice

**Surendra Singh,[1] J. Xiong,[2] A. P. Chen,[2] M R Fitzsimmons,[3] Q. X. Jia[2]**

[1]Solid State Physics Division, Bhabha Atomic Research Centre, Mumbai 400094 India

[2]Center for Integrated Nanotechnologies, Los Alamos National Laboratory, Los Alamos, NM, USA

[3]Quantum Condensed Matter Division, Oak Ridge National Laboratory, Oak Ridge, TN, USA

We report the observation of field-induced magnetization of $BiFeO_3$ (BFO) in an ultrathin BFO/$La_{0.7}Sr_{0.3}MnO_3$ (LSMO) superlattice using polarized neutron reflectivity (PNR). Our PNR results indicate parallel alignment of magnetization across BFO/LSMO interfaces. The study showed an increase in average magnetization on increasing applied magnetic field at 10K. We observed a saturation magnetization of 110±15 kA/m (~0.8 $\mu_B$/Fe) for ultrathin BFO layer (~2 unit cell) sandwiched between ultrathin LSMO layers (~ 2 unit cell), which is much higher than the canted moment (0.03 $\mu_B$/Fe) in the bulk BFO. The macroscopic VSM results on superlattice clearly indicate superparamagnetic behavior typically observed in nanoparticles of manganites.

**Introduction:**

In complex oxides the competition between, charge, spin, orbital, and lattice degrees of freedom lead to fascinating functional behavior. Due to breaking of space and/or time symmetries intrinsic to an interface, additional unique properties can emerge that are absent in bulk materials,.[1,2,3,4] Specifically, coupling between ferromagnetic (FM), antiferromagnetic (AFM), and ferroelectric (FE) order parameters via strain or magnetoelectric effects may yield new routes to achieve strong magnetoelectric coupling in a multiferroic.[2,5,6]

Being a propotypical multiferroic, $BiFeO_3$ (BFO) has received much scrutiny.[7,8,9] BFO is a single phase multiferroic material which exhibits magnetoelectric coupling between FE ($T_C$ = 1103 K) and AFM ($T_N$ = 643 K) order parameters. There have been mixed reports about the origin of ferromagnetism in BFO films,[7,8,9,10] which include epitaxial strain,[7,9] oxygen deficiency or presence of $Fe^{2+}$,[8] and deformation of oxygen octahedral.[10] Indeed, theoretical calculations suggest a weak ferromagnetism in BFO can result from canting of the AFM structure due to Dzyaloshinskii-Moriya interaction; however, the estimated value of uncompensated magnetization can be much smaller than observed.[11]

In addition to interest in BFO films, there has also been interest, motivated by the possibility to realize magnetoelectric coupling, in heterostructures of BFO and $La_{0.7}Sr_{0.3}MnO_3$ (LSMO).[12,13,14] Magnetic moments at the interface of BFO and LSMO hetrostructure, reported earlier, have been attributed to Fe-Mn hybridization and orbital reconstruction, which is associated with charge transfer possibly occurring in close proximity to the BFO/LSMO interface.[12, 15]

Previous experimental works[12,16,17] clearly found an induced ferromagnetism at the BFO/LSMO interface. Yu *et al.*,[12] indicated an interface magnetization of 0.6 $\mu_B$/Fe, which are ferromagnetically coupled to the LSMO layer. Whereas the polarized neutron reflectivity (PNR) measurements on BFO/LSMO hetrostructures suggested a similar magnetization for whole BFO layer (~ 6 u. c., thick) but with different magnetic couplings: ferromagnetic[16] and antiferromagnetic.[17] The different coupling may be due to different surface termination of the surface of LSMO in these hetrostructures. However growth of LSMO layer with different termination ($MnO_2$ and SrO) depends on many parameters[18] e.g. nature of substrate (polar or non polar) etc., and it is difficult to measure it experimentally for such ultrathin hetrostructures. Further, *ab initio* calculations also confirmed both AFM and FM exchange coupling across the BFO/LSMO interface strongly depends on the termination of the LSMO film.[16] Theoretical calculations also suggested such magnetization can indeed develop via spin-exchange effects in a BFO/LSMO superlattice, which results to strong induced magnetization for even thinner films (~3 u. c.).[16] However it is important to see the variation of induced magnetization of ultrathin BFO layer in BFO/LSMO hetrostructure as a function of magnetization of LSMO layer. In view of this we investigated the magnetization depth profile across an ultrathin BFO/LSMO superlattice (with 2 u. c. thick BFO layers) as a function of applied magnetic field along the hysteresis curve at low temperature.

**Experimental:**

A superlattice structure of $[(BFO)_m/(LSMO)_n]_N$ was grown on an (001) STO substrate by pulsed laser (KrF) deposition, where the *m* and *n* are 2 u. c. of BFO and LSMO, respectively, and *N* = 32 is the number of repetition of BFO/LSMO bilayer. The deposition rate was controlled through appropriate focus of laser beam on the target, where the nominal growth rate was ~0.01Å/pulse. The substrate temperature during film growth was initially optimized and was maintained at 750 °C as calibrated by the pyrometer. The oxygen pressure during deposition was 200 mTorr. The present sample and the thicker hetrostructure studied earlier[16] were grown in identical condition. Evidence of structurally well-defined interfaces of similar hetrostructures was obtained using x-ray diffraction, transmission electron spectroscopy (TEM) and Rutherford Backscattering Spectroscopy.[16] The magnetization hysteresis of the superlattice was measured with vibrating sample magnetometer (VSM). Depth dependent structure and magnetization of the superlattice was analyzed by X-ray reflectivity (XRR) and polarized neutron reflectivity (PNR) techniques.[16,19,20,21,22] X-ray diffraction (XRD) (Fig. 1a) and XRR (Fig. 1b) were measured using Cu $K_\alpha$ radiation and PNR was measured at Asterix spectrometer at Los Alamos Neutron Scattering Center (LANSCE).[19]

In case of XRR and PNR the reflected radiation is measured from a sample as a function of wave vector transfer [$Q = 4\pi \sin(\theta)/\lambda$] perpendicular to the sample surface where $\lambda$ and $\theta$ are x-ray or neutron wavelength and angle of incidence respectively. The specular reflectivity, $R$, is determined by the scattering length density (SLD) depth profile, $\rho(z)$.[19,20,21] For XRR, $\rho(z)$ is proportional to electron density whereas for PNR, $\rho(z)$ consists of nuclear and magnetic SLDs such that $\rho^\pm(z) = \rho_n(z) \pm CM(z)$, where $C = 2.91 \times 10^{-9}$ Å$^{-2}$ cm$^3$/emu and $M(z)$ is the magnetization (a moment density obtained in emu/cm$^3$) depth profile.[19] The +(-) sign denotes neutron beam polarization parallel (anti-parallel) to the applied field and corresponds to reflectivities, $R^\pm(Q)$. The layer structure were obtained from the XRR data by fitting model ESLD profiles, $\rho(z)$ that fit the reflectivity data. The reflectivities were calculated using the dynamical formalism of Parratt[23] and parameters of the model were adjusted to minimize the value of reduced $\chi^2$ –a weighted measure for goodness of fit.[24] Layers in a model consisted of regions with different electron SLDs. The parameters of a model included layer thickness, interface (or surface) roughness and electron SLD.

**Results and Discussion:**

Fig. 1(a) presents a θ-2θ XRD pattern of BFO/LSMO heterostructure. It is evident from this pattern that all the layers show (00l) texture. Together with the in-plane XRD scan (not shown here), the experimental results confirm epitaxial growth across the hetrostructure. Inset (i) shows the in large version of XRD peak around STO (001) reflection. Inset (ii) shows the rocking curve of the BFO (002) peak, where the full width at half-maximum (FWHM) is about 0.09°, suggesting high crystalline quality of the superlattice.

Fig. 1 (b) shows the XRR data (closed circles) normalized to the asymptotic value of the Fresnel decay ($R_F = \frac{16\pi^2}{Q^4}$ )[19] as a function of wave vector transfer ($Q$), from the sample. Inset of Fig. 1 (c) shows the electron scattering length density (ESLD) depth profile of sample which gave best fit (solid line) to XRR data. The ESLD profile confirms the periodic multilayer growth. We obtained a bilayer thickness (thickness of LSMO layer + thickness of BFO layer) of ~12.7±1.5 Å, from XRR results. The interface averaged root-mean-square roughness for BFO/LSMO (BFO on LSMO) and LSMO/BFO (LSMO on BFO) interface obtained from XRR is 1.6 ± 0.5 Å and 0.7 ± 0.3 Å, respectively.

The macroscopic magnetic characteristics of the hetrostructure were investigated as a function of temperature and magnetic field. Fig. 2 (a) shows the magnetization vs. magnetic field hysteresis along the plane at 10 K from the superlattice. The saturation magnetization of about 160 kA/m was achieved at a field of ~6 kOe. Closed circles (blue) on M-H plot are the fields on which we have made PNR measurements and are discussed later. Fig. 2 (b) shows VSM data for the sample after conditions of field-cooling (FC) (cooling field $H_{FC}$ = 1 kOe) and zero-field-cooling (ZFC). The FC data were collected during the warming cycle in a field of 1 kOe (▲) and 0 Oe (○). ZFC data were collected during the warming cycle in a field of 1 kOe. We observed a strong separation of the FC and ZFC measurements at the blocking temperature of ~45 K when measured in an applied field of 1 kOe as the sample was warmed. In addition, the thermoremanent magnetization of the FC sample falls sharply to minimum value at the blocking temperature. The behavior of the magnetization of our ultrathin layers comprising the superlattice is reminiscent of superparamagnetic behavior typically observed in nanoparticles of manganites.[25] At the blocking temperature (~ 45 K) the ZFC data show a peak in magnetization, which coincides with the loss of thermoremanent magnetization measured in FC condition (in an applied field of 0 Oe).

It is evident from Fig. 2 (a) that FC loop is shifted towards the negative field value. The shift may be quantified through the exchange field parameter[26]: $H_{ex}$ = -($H_{right}$ + $H_{left}$)/2, whereas the coercivity is defined as $H_C$ = ($H_{right}$ - $H_{left}$)/2, $H_{right}$ (≈ 360 Oe) and $H_{left}$ (≈ -385 Oe) being the points where the loop intersects the field axis. We obtained coercive fields ($H_c$) of about 372 Oe, and exchange bias ($H_{ex}$) of about -12 Oe at 10 K. This shift in exchange field (~ -1.2 mT) is very small (~ 1/5$^{th}$) then that observed by Wu et al.[14] However similar shift (~ 2 mT) were observed in thicker superlattice by Jain et al.[17]

PNR measurements were performed at different temperatures while warming the sample in a field of 1 kOe after cooling in a field of 1 kOe (FC) from 300K. PNR involves specular reflection of polarized neutron from magnetic film as a function of $Q$.[16,19,20] Specular reflection of the neutron beam with polarization parallel (+) and anti-parallel (-) to sample magnetization corresponds to reflectivities, $R^{\pm}(Q)$. Fig. 3 (a) shows the $R^{\pm}(Q)/R_F$ for 300 K and 10 K. Fig. 3 (b) and (c) show the normalized spin asymmetry (NSA) defined as $(R^+ - R^-)/R_F$ at these temperatures which highlights the spin dependence of the reflectivity. The amplitude and period of the oscillatory variation in NSA are related to the magnetization contrast across interfaces between magnetic/non-magnetic layers and the total thickness of the film respectively.[21]

To extract the magnetization profile from the PNR measurements we fixed the parameters (thickness of layers, interface roughness and number density of each layer) obtained from XRR data and only the magnetization depth profile was optimized. PNR data at 300 K show negligible spin dependent (nearly zero NSA data) reflectivity suggesting very small magnetization ~ 20 ± 15 kA/m for LSMO layer and zero magnetization for BFO layer. While PNR data at 10 K (FC condition) in a magnetic field of 1kOe (higher than $H_c$) clearly show spin dependent neutron reflectivity [with large amplitude NSA data in Fig. 3 (c)]. Fig. 3 (d) and (e) show the nuclear scattering length density (NSLD) and magnetization depth profiles, respectively, which yield the continuous lines in Fig. 3 (a-c)). We obtained a magnetization of 100 ± 15 kA/m and 29 ± 12 kA/m for LSMO and BFO layers, respectively. The thickness averaged magnetization for superlattice was 72 ± 12 kA/m, which is comparable to the average magnetization of ~ 80 kA/m obtained from vibration sample magnetometry (VSM) (Fig. 2 (b)) under the same measurement conditions.

In order to see the effect of induced magnetization on BFO as a function of the magnetization of LSMO layer (LSMO being ferromagnetic will show increase in average magnetization on application of magnetic field till it get saturated) we measured PNR data as a function of applied magnetic field (~ 1 kOe to 6.1 kOe). The VSM data also suggest the superlattice magnetization is still approaching saturation at very high fields of 6 kOe. Fig. 4 (a) – (d) shows the PNR (NSA) data at different applied magnetic field. Spin dependent PNR data clearly depicts an increase in amplitude of NSA profile on increasing the magnetic field which indicates an increase in average magnetization on increasing magnetic field. Fig. 4 (f) shows the magnetization profiles across a bilayer of LSMO/BFO which gave best fit to NSA data (solid lines in (a) – (d)) at successively higher fields.

For fitting PNR data we have also assumed different magnetization profiles for BFO layer. Fig. 4 (d) shows the fits to NSA data at 4.5 kOe at 10 K, assuming zero magnetization of BFO layer (open star, blue) and magnetization of BFO layer antiparallel (antiferromagnetically coupled) to that of LSMO layer (open triangle, green) along with best fit (solid line) when both are parallel (ferromagnetically coupled). It is clear from Fig. 4 (d) that magnetization of BFO and LSMO layers are ferromagnetically coupled.

Average magnetization of LSMO layer extracted from PNR measurements (Fig. 5) increases from $100 \pm 15$ kA/m to $218 \pm 22$ kA/m on increasing magnetic field from 1 kOe to 4.5 kOe at 10 K under FC condition. On further increasing magnetic field to 6.1 kOe we observed a small change in the magnetization (~ $223 \pm 21$ kA/m) of the LSMO layer, suggesting saturation of LSMO layer. For the BFO layer, we obtained an increase in the average magnetization from $28 \pm 10$ kA/m to $109 \pm 15$ kA/m while increasing the magnetic field from 1 kOe to 4.5 kOe at 10 K under the FC condition. The magnetization of the BFO layer remained nearly unchanged (~ $112 \pm 15$ kA/m) as the field was increased to 6.1 kOe. The field dependence of the BFO magnetization is the same as that of the LSMO magnetization (Fig. 5). Previously, the temperature dependencies of the BFO and LSMO magnetizations in another BFO/LSMO superlattice were also found to be the same.[17] These results are consistent with the hypothesis that the LSMO magnetization induces magnetization in the BFO.

Using PNR we have also estimated the thickness-weighted average magnetization of whole superlattice at different magnetic field and shown in Fig. 5 (open squares, black). These magnetizations of superlattice obtained from PNR are consistent with the average/saturation magnetization of the superlattice measured by the macroscopic VSM technique at different magnetic fields of 1.0 kOe, 4.5 kOe and 6 kOe (open star, in Fig. 5).

Another important point is the behavior of $M$(T) data measured by macroscopic VSM technique for FC and ZFC condition which indicate superparamagnetic behavior of manganite nanoparticles. Thus such properties may be an added interaction at interface, in addition to features associated with the superlattice,[17] *e.g.,* growth, strain, proximity to a ferromagnet, *etc* which results for such a large magnetization and magnetic coupling in BFO/LSMO superlattice. Further we observed a minimum magnetic moment of ~0.8 $\mu_B$/Fe for BFO layer at low temperature which is higher than ~0.6 $\mu_B$/Fe at BFO/LSMO interface observed by Yu *et al.*,[12] which they have attributed to possible electronic orbital reconstruction at the interface.

**Conclusions:**

In summary we have measured the magnetization depth profile of ultrathin BFO/LSMO superlattice as a function of applied in-plane magnetic field at 10 K under FC condition. We found that the field dependence of the magnetizations LSMO and BFO layers (thickness ~ 2 u. c for each) are the same, which suggests an intimate relationship between the LSMO and BFO uncompensated magnetization order parameters (previously only the temperature dependence of the magnetizations had been reported). Macroscopic VSM magnetization measurement also suggested superparamagnetic behavior of manganite nanoparticles, which may be responsible for added coupling of magnetic properties across BFO and LSMO. Our comprehensive magnetization depth profiling using PNR data imply an enhancement in saturation magnetization (~ 110 ± 15 kA/m or ~ 0.8 $\mu_B$/Fe) of ultrathin BFO layer at 10 K when BFO is placed in contact with ferromagnetic LSMO.

**Acknowledgements:**

This work was supported by the U.S. Department of Energy through the LANL/LDRD program and was performed, in part, at the Center for Integrated Nanotechnologies (CINT), a U.S. Department of Energy, Office of Basic Energy Sciences user facility. Los Alamos National Laboratory, an affirmative action equal opportunity employer, is operated by Los Alamos National Security, LLC, for the National Nuclear Security Administration of the U.S. Department of Energy under contract DE-AC52-06NA25396. This research partially supported by the Laboratory Directed Research and Development Program of Oak Ridge National Laboratory, managed by UT-Battelle, LLC, for the U. S. Department of Energy.

**Figure Captions**

Fig. 1 (a) Typical θ-2θ (out of plane) XRD pattern (in log axis) of the superlattice [(LSMO)$_2$/(BFO)$_2$]$_{32}$ on (001) SrTiO$_3$ substrate showing high quality, epitaxial multilayer. Inset (i) shows larger version of data around (001) STO peak. Inset (ii) shows the rocking curve with FWHM of $0.0923^0$ of BFO (001) diffraction peak. (b) X-ray reflectivity (XRR) from superlattice. Inset shows the depth profile of the electron scattering length density (ESLD) of the superlattice which gives the best fit to XRR data.

Fig. 2 (a) Magnetic hysteresis curve at 10 K of the superlattice [(LSMO)$_2$/(BFO)$_2$]$_{32}$ on (100) SrTiO$_3$ substrate. (b) Magnetization as a function of temperature for field cooled (FC) at 1kOe and zero field cooled (ZFC) condition on warming in different magnetic fields.

Fig. 3 (a) Polarized neutron reflectivity (PNR) from superlattice [(LSMO)$_2$/(BFO)$_6$]$_2$ on (100) SrTiO$_3$ substrate at 300 K and 10 K under field cooled (FC) condition in a field of 1 kOe. Reflectivity data at 300 K and 10 K are shifted by a factor of 5for the sake of clarity. Normalized spin asymmetry (NSA), data at 300 K (b) and 10 K (c). (d) Nuclear scattering length density (SLD) and (e) magnetization (*M*) depth profile extracted from fitting PNR data at 300 K and 10 K.

Fig. 4 (a)-(d) Normalized spin asymetry (NSA), ($R^+ - R^-$)/$R_F$, data (PNR data) at different applied in-plane magnetic field. (e) NSA data at 4.5 kOe with fits assuming different magnetization depth profiles. (f) Magnetization depth profiles across a bilayer of LSMO/BFO at different magnetic field which gave best fit to NSA data [(a)-(d)].

Fig. 5 Average magnetization of LSMO and BFO layer as a function of applied in-plane magnetic field obtained from PNR. Thickness weighted magnetization and average

magnetization for whole superlattice obtained from PNR (open square, black) and VSM (open star), respectively as a function of magnetic field.

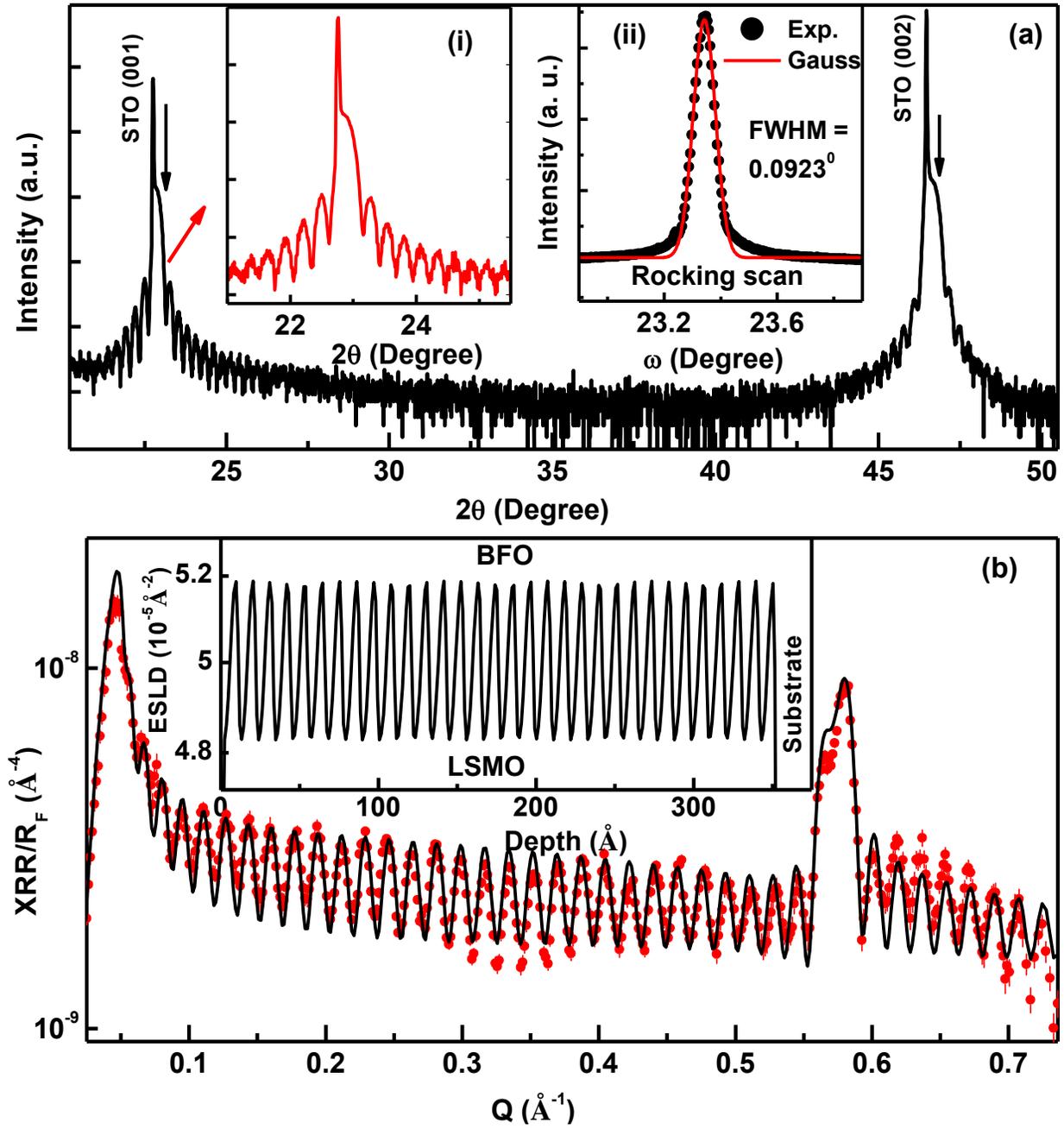

Fig. 1

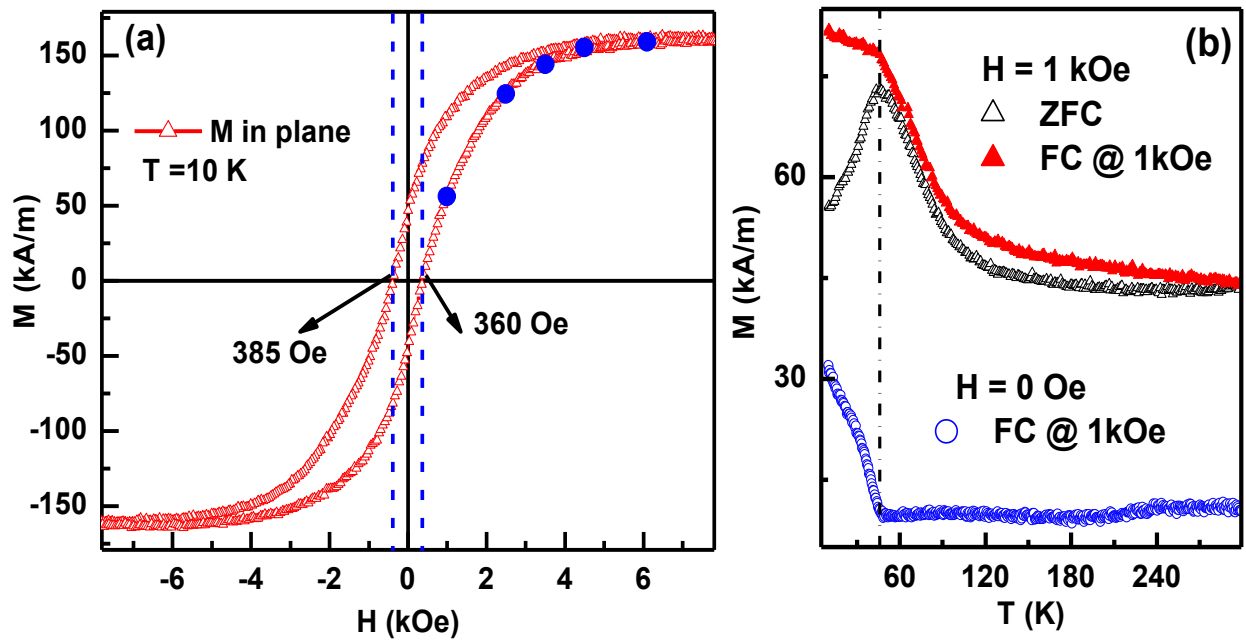

Fig. 2

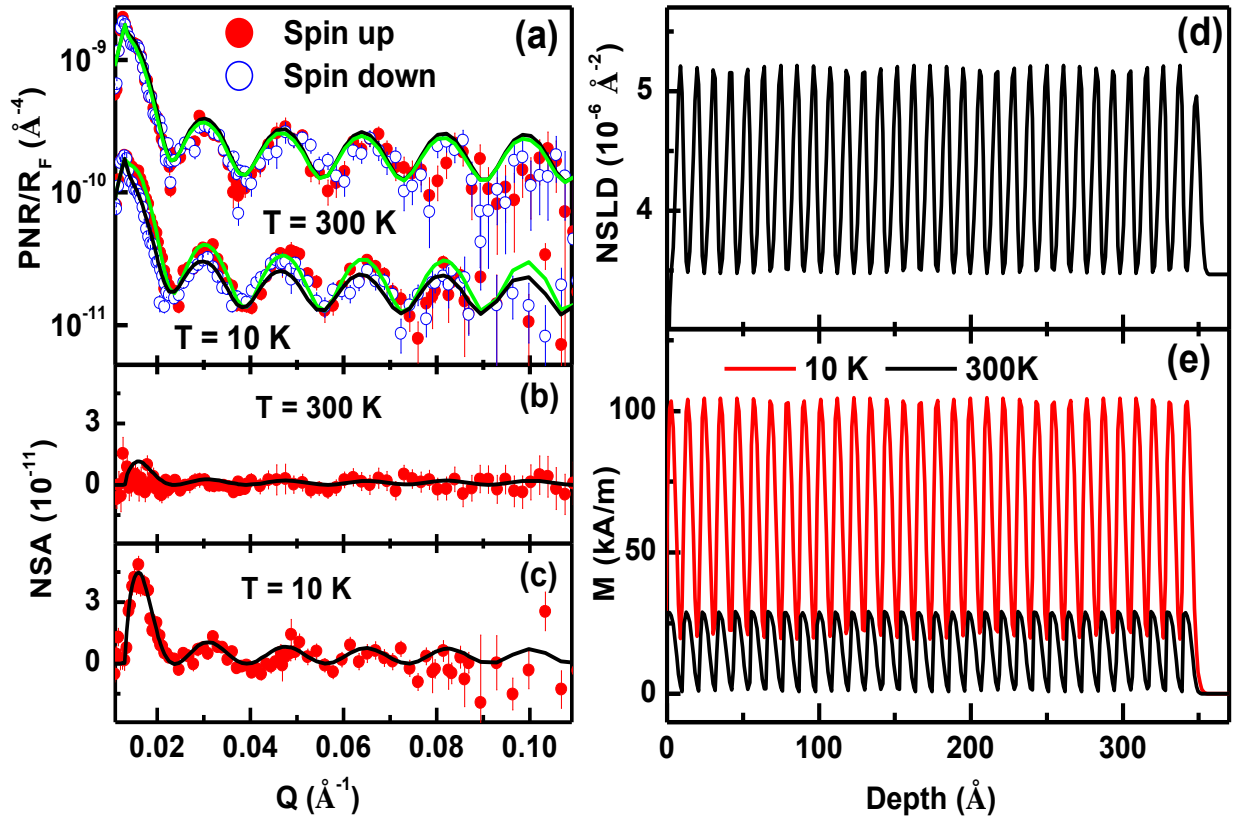

Fig. 3

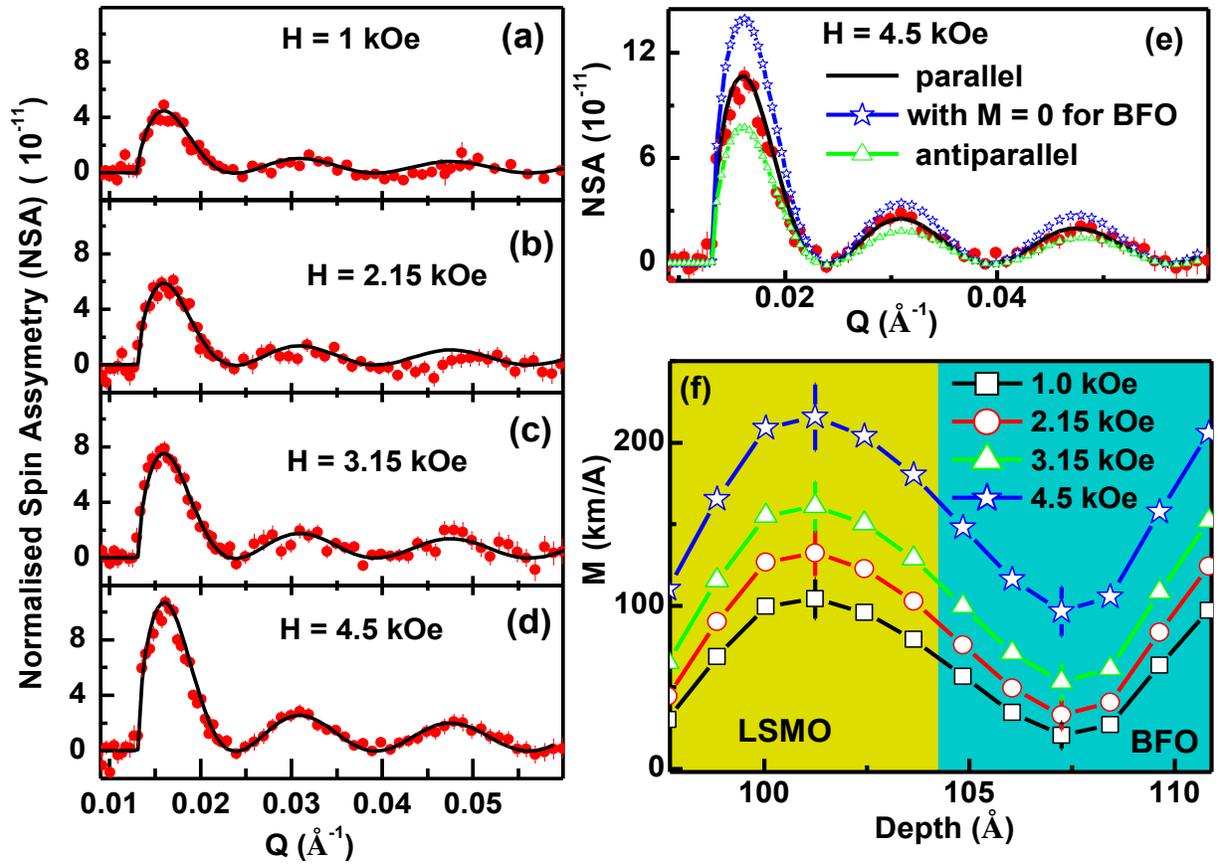

Fig. 4

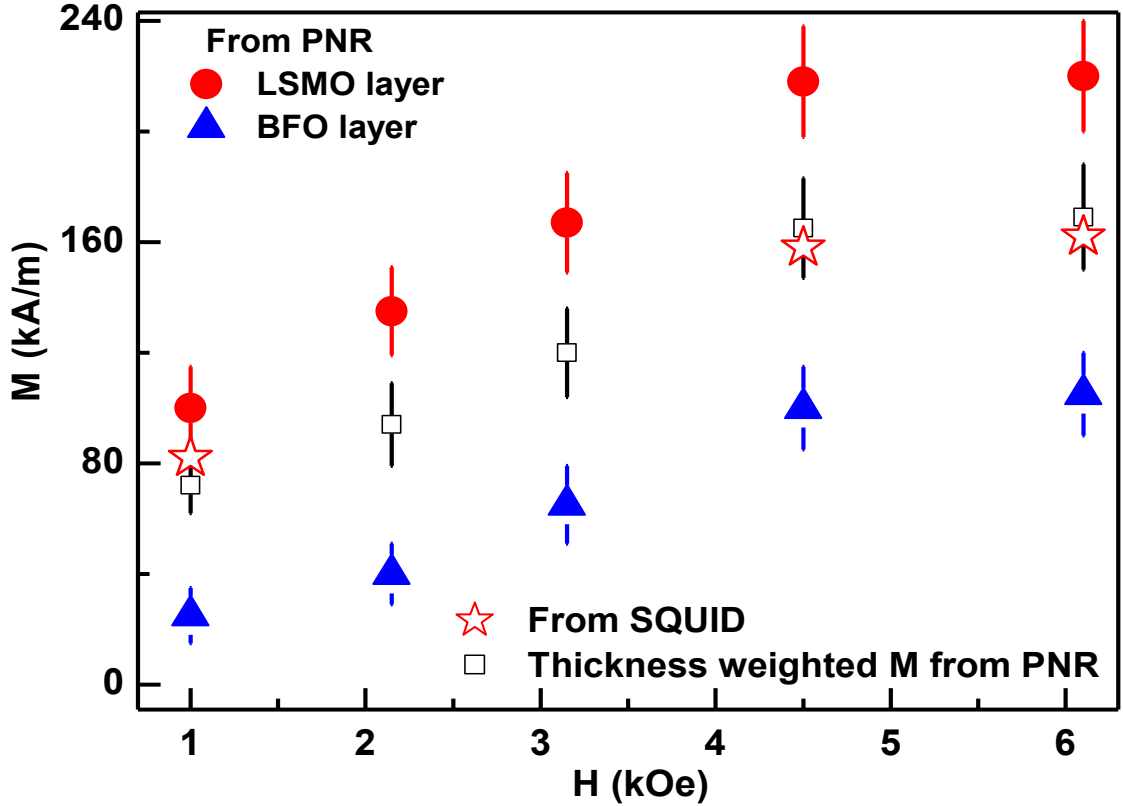

Fig. 5